\documentclass{INTERSPEECH2023}

 \interspeechcameraready

\usepackage[utf8]{inputenc} 
\usepackage[T1]{fontenc}    
\usepackage{hyperref}       
\usepackage{url}            
\usepackage{booktabs}       
\usepackage{amsfonts}       
\usepackage{nicefrac}       
\usepackage{microtype}      
\usepackage{xcolor}         
\usepackage{graphicx}
\usepackage{multicol}
\usepackage{multirow}
\usepackage{color, colortbl}
\usepackage{amsbsy}
\usepackage{array, makecell} %
\usepackage{boldline} 
\usepackage{hhline}
\usepackage[numbib]{tocbibind}
\usepackage[numbers,sort&compress]{natbib}
\usepackage[labelfont=bf]{caption}
\usepackage{todonotes}
\newcolumntype{P}[1]{>{\centering\arraybackslash}p{#1}}
\newcolumntype{M}[1]{>{\centering\arraybackslash}m{#1}}
\newcolumntype{N}{@{}m{0pt}@{}}

\definecolor{LightGray}{gray}{0.9}

\title{Lexical Speaker Error Correction: Leveraging Language Models for Speaker Diarization Error Correction}

\name{Rohit Paturi$^*$, Sundararajan Srinivasan$^*$, Xiang Li}
\address{AWS AI Labs}
\email{paturi@amazon.com, sundarsr@amazon.com, xiangzai@amazon.co.uk}

\begin{document}

\maketitle
\def\thefootnote{*}\footnotetext{These authors contributed equally to this work}\def\thefootnote{\arabic{footnote}}

\begin{abstract}
Speaker diarization (SD) is typically used with an automatic speech recognition (ASR) system to ascribe speaker labels to recognized words. The conventional approach reconciles outputs from independently optimized ASR and SD systems, where the SD system typically uses only acoustic information to identify the speakers in the audio stream. This approach can lead to speaker errors especially around speaker turns and regions of speaker overlap. In this paper, we propose a novel second-pass speaker error correction system using lexical information, leveraging the power of modern language models (LMs). Our experiments across multiple telephony datasets show that our approach is both effective and robust. Training and tuning only on the Fisher dataset, this error correction approach leads to relative word-level diarization error rate (WDER) reductions of 15-30\% on three telephony datasets: RT03-CTS, Callhome American English and held-out portions of Fisher.

\end{abstract}

\section{Introduction}

\begin{figure*}[htp]
  \centering
  \includegraphics[width=1\linewidth]{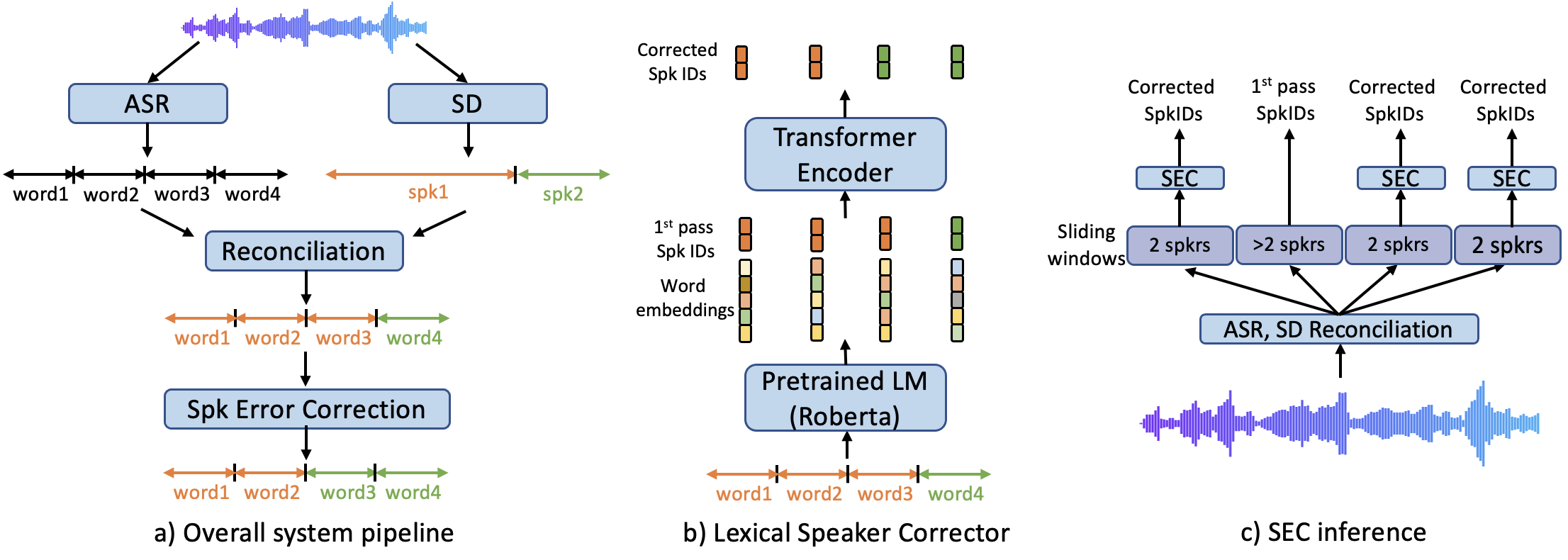}
  \vspace{-\baselineskip}
  \caption{(a) Speaker Error Correction as a 2nd-pass post-processing step to the traditional SD-ASR system, (b) Lexical SEC: Word embeddings from the LM, Speaker IDs from SD are passed to the Transformer Encoder to get the corrected Speaker IDs, (c) SEC inference performed on sliding windows with 2 hypothesis speakers.}
  \vspace{-1\baselineskip}
\end{figure*}

Speech transcription systems have advanced significantly in the past decade but even with these remarkable advances, machines have difficulties understanding natural conversations with multiple speakers such as in broadcast interviews, meetings, telephone calls, videos or medical recordings. One of the first steps in understanding natural conversations is to recognize the words spoken and their corresponding speakers. Speaker Diarization (SD) is the process of determining "who spoke when" in a multi-speaker audio signal and is a key component in any speech transcription system. SD is used in conjunction with Automatic Speech Recognition (ASR) to assign a speaker label to each transcribed speaker turn and has widespread applications in generating meeting/interview transcripts, medical notes, automated subtitling and dubbing, downstream speaker analytics, among others (we refer to this combined system as SD-ASR in this paper). This is typically performed in multiple steps that include (1) transcribing the words using an ASR system, (2) predicting “who spoke when” using a speaker diarization (SD) system, and, finally, (3) reconciling the output of those two systems.

Recent advances in SD systems are outlined in \cite{park2022review} and the independent module optimized SD systems typically consists of the following main sub-tasks: (a) segment the input audio into speech segments using a Voice activity detector (VAD), (b) generate speaker segments  from the speech segments by either using a uniform window size \cite{32wang2018speaker,zhang2019fully,garcia2017speaker} or by detecting speaker turns \cite{yin2018neural,park2018multimodal,xia2022turn}, (c) extract speaker embeddings \cite{li2017deep,snyder2018x,32wang2018speaker,dawalatabad2021ecapa} for each of the speaker segments and (d) cluster the resulting speaker embeddings using clustering algorithms like Spectral Clustering \cite{32wang2018speaker}, Agglomerative Hierarchical Clustering \cite{garcia2017speaker} among others. These sub-tasks of most of the diarization systems in literature rely only acoustic information and can thus lead to speaker errors, mainly around the speaker turns. This can happen in uniform speaker segmentation as long segments very likely contain speaker turn boundaries, while short segments carry insufficient speaker information. It is also shown that detecting speaker turns using only acoustic information is also error-prone \cite{xia2022turn}. In addition to the SD errors, speakers can be attributed to the wrong words in the SD-ASR reconciliation phase due to errors in ASR word timings. Reconciliation errors can also occur in regions of speech overlap as SD can identify one of the speakers while ASR can identify words corresponding to a different speaker.

Lexical information can contain complementary information which can be very useful in accurately predicting speaker turns \cite{park2018multimodal,xia2022turn}. For instance, analyzing only the written transcript of a conversation such as "how are you i am good", enables us to infer that there is likely a speaker change between the utterances "how are you" and "i am good". There have been a handful of works \cite{park2018multimodal,xia2022turn, park2020speaker,shafey2019joint,india2023language,flemotomos2019linguistically} which leverage the ASR transcripts to infuse lexical information in the SD module. In [7], lexical cues are used to estimate the speaker turns for diarization. \cite{park2020speaker} made use of turn probabilities from lexical cues in the clustering stage by enhancing the adjacency matrix. Though these approaches showed good SD improvements, these systems can still produce errors around speaker turns due to ASR and Diarization errors in overlapped speech as well are sensitive to ASR word timings as they rely on ASR timings in the diarization sub-tasks as well as in the Reconciliation phase. \cite{shafey2019joint} modeled SD and ASR jointly but is confined to 2 speakers with specific distinct roles.

In this paper, we propose a Speaker Error Correction (SEC) module which can correct speaker errors at the word level without modifying the underlying ASR or the acoustic SD system. This SEC module makes use of the any of the readily available pre-trained LMs \cite{devlin2018bert,liu2019roberta,yang2019xlnet,brown2020language} to infuse the lexical knowledge to correct speaker errors while also leveraging speaker scores from the SD system to prevent over-corrections. The reliance on LMs also significantly reduces the amount of speaker labelled text data needed to train the system. Our approach has components which are modular and don’t need paired audio, text data to train while only needing a small amount of paired data for fine-tuning. This approach is also easier to integrate with existing systems than other lexical-based diarization approaches, since the first-pass acoustic SD system can be run independently of the ASR system. Using experiments across three telephony datasets, we demonstrate that the proposed system is both effective as well as capable of generalization.
\section{Speaker Error Corrector}
\label{SS}

The overall pipeline of the proposed two-pass Speaker Error Corrector (SEC) framework is shown in Fig 1a. The conventional Speaker Transcription system consists of an ASR module, a SD module and a reconciliation stage. The SEC follows the reconciliation stage and takes in two streams of inputs: acoustic features from the SD module and lexical features from the ASR module. The ASR and acoustic SD models can continue to run in parallel, making it easier to integrate with existing systems. The core component of the SEC is the Lexical Correction module which takes in the transcribed words from ASR along with the speaker labels from the SD module. These are explained in more detail in the following sub-sections. 
\subsection{Lexical Diarization Corrector}
While lexical features have complementary information to the acoustic features and can be leveraged to correct some of the errors from a naïve reconciliation of ASR and SD, lexical features alone can’t accurately predict the speaker labels especially in realistic conversations. So, we propose a simple yet efficient way to correct the speakers based on both the decisions from the 1st pass diarizer and the ASR transcriptions.
Our proposed Lexical Speaker Error Corrector consists of two main components:  a backbone language model (LM) and a Transformer Encoder Front-end to predict the speaker labels. After reconciling ASR and diarization outputs, we have speaker labels \(\{\boldsymbol{S_i}\}_{i=1}^N, \boldsymbol{S_i} \in \mathbb{R}^{1\times K}\) for every word \(\{\boldsymbol{W_i}\}_{i=1}^N\), where \(N\) is the number of words in the sequence and K is the number of speakers the SEC is trained to handle. The words \(W_i\) are tokenized and passed to the backbone LM to obtain contextual word embeddings \(\{\boldsymbol{E_j}\}_{j=1}^M, \boldsymbol{E_j} \in \mathbb{R}^{1\times W}\) where \(M\) is the number of tokens in the word sequence and W is the word embedding dimension. The word level speaker labels \(S_i\) are mapped to token level by mapping the speaker ID corresponding to the word to its first token if the word has more than 2 tokens and assigning a special “don’t care” token to any of the subsequent tokens of the word. These token level embeddings \(E_j\) are concatenated with the speaker IDs \(S_j\) to form the fused features for the Front-end Transformer Encoder as shown in Figure 1b. The posteriors from the Front-end Encoder \(\{\boldsymbol{L_{ij}}\}_{j=1}^K, \boldsymbol{L_i} \in \mathbb{R}\) are used to optimize the classification loss on the ground-truth speaker labels.

\subsection{Training Methodology}
The SEC model can be trained only using speaker turn transcripts and doesn't require paired audio data and we show that training the lexical corrector on just the transcripts also improves the performance of the baseline. Since the relatively smaller number of speaker errors produced by 1st pass diarizer system limits the training of the error corrector, we train the corrector by simulating speaker errors based on the ground truth as well by simulating ASR substitution errors. 

We define the probability of ASR errors as \(P_{ASR}\) and the probability of speaker errors as \(P_{Spk}\). Setting \(P_{ASR}=1\) implies that all the words in the training transcripts are substituted with random words and \(P_{ASR}=0\) implies the original ground-truth transcripts. Similarly,  \(P_{Spk}=1\) implies all the speaker labels are randomly substituted whereas \(P_{Spk}=0\) implies the ground-truth speaker labels. We simulate ASR, Speaker errors using a curriculum learning paradigm \cite{bengio2009curriculum} to make sure that we don’t under or over correct the speakers and balance the information flow from the SD labels and ASR word lexical information. We start the curriculum for  \(P_{Spk}\) at a low value and increase \(P_{Spk}\) as the training progresses. Conversely, \(P_{ASR}\) starts at a high value at the first epoch and decreases as the training progress. The intuition for this curriculum with \(P_{ASR}\) being higher and \(P_{Spk}\) being lower in the initial epochs is to train the model without any meaningful lexical information and to train the model to at least copy the 1st pass speaker labels in the initial epochs. More meaningful lexical information with a smaller \(P_{ASR}\) is used in the later epochs along with a higher \(P_{Spk}\) to train the model on more complex speaker errors as the training progresses. 

In addition to the errors simulated text data, we also use paired audio data to train, fine-tune the model on real data. For this, we generate speaker labels using the baseline 1st SD and use the ground-truth speaker labels as the targets. In this work, we train the SEC on two speaker cases, i.e., \(K=2\)

\subsection{Inference Setup}
During inference, we perform error correction on sliding windows with a fixed number of ASR transcribed words as shown in Fig 1c. Though the lexical corrector is trained to only correct two speakers locally, we can still handle use-cases where more than two speakers are detected globally in the audio. We achieve this by only correcting sliding windows comprising of two speakers and by bypassing the remaining windows as shown in Figure 1c. The size of the sliding window is a parameter we tune on a validation set.

\section{Experiments}
\label{exps}

\subsection{Data and Metrics}
In this work, we use the full Fisher dataset \cite{40LDC, 41LDC} to train the Speaker Corrector system. We split the Fisher data into train, validation and test splits as defined in \cite{wang2022highly}. We also the Fisher train set to fine-tune the backbone LM model as well as to train, fine-tune the Corrector model. We only use the Fisher validation split for tuning our model. For evaluation, in addition to Fisher test split, we use the standard dev, test splits of CALLHOME American English (CHAE) \cite{39LDC} and RT03-CTS \cite{42LDC} which are majorly two speaker calls. We also evaluate on the two-speaker only set of CHAE, the CH-109 dataset \cite{cyrta2017speaker} by fixing the number of clusters to 2 as well as automatically determining the number of speakers in the 1st pass SD system.

In order to evaluate the full ASR, SD system, we use the Word Diarization Error Rate (WDER) proposed in \cite{shafey2019joint} as it aptly captures both ASR and SD errors at the word level. We also account for words transcribed in regions of speech overlap in the WDER metric. This is achieved by using asclite \cite{fiscus2006multiple} as it can align multiple speaker hypotheses against multiple reference speaker transcriptions and can also efficiently handle words in regions of speaker overlaps.

\begin{figure}[t]
  \centering
  \includegraphics[width=1\columnwidth]{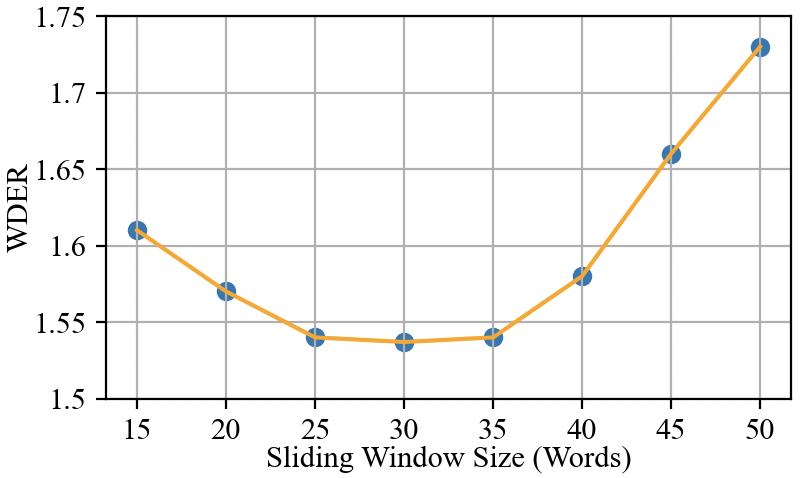}
  \vspace{-1\baselineskip}
  \caption{Window Size tuning on the validation set}
  \label{fig:Window Tuning}
  \vspace{-1\baselineskip}
\end{figure}

\subsection{Baseline System}
Our baseline SD system follows the pipeline in \cite{32wang2018speaker} and consists of a speaker embedding model followed by Spectral Clustering and the number of speakers is identified using the maximum eigengap of the Spectral Clustering. The speaker embedding model is based on a ResNet-34 architecture trained with a combination of classification, metric loss \cite{chung2020defence} and channel loss \cite{higuchi2020speaker} on about 12k speakers and 4k hours of CTS data . We use a uniform speaker segmentation \cite{zhang2019fully,garcia2017speaker} with a duration of 500ms to extract the speaker embeddings followed by the Clustering phase for the SD system. Our baseline SD system is comparable to state-of-the-art diarization systems across several datasets and achieves a a DER of 3.72 and SER of 1.1  on CHAE test set which is a stronger baseline than the one reported in \cite{park2020speaker}. We use a hybrid ASR system \cite{zhou2020rwth, wang2020transformer,yang2022conformer} with a Conformer Acoustic model \cite{gulati2020conformer} and a n-gram Language model trained on several tens of thousands of audio, text data. For the reconciliation phase, the SD system provides speaker turns with time boundaries and these labels are mapped to recognized words using the associated word boundaries from the ASR system. When the speaker turn boundary falls in the middle of a word, we assign the word to the speaker with the largest overlap with the word similar to the baseline system in \cite{shafey2019joint}. We use a neural-network based Speech Activity detector (SAD) similar to \cite{majumdar2020matchboxnet} as a front end for both SD-ASR systems above.

\subsection{SEC System}
For the SEC model, we use a pre-trained Roberta-base model \cite{liu2019roberta} as the backbone LM and a Transformer Encoder of size 128 hidden states for the Front-end model. The curriculum for \(P_{ASR}\) starts at 1 at the 1st epoch and decreases to 0.08 in the 10th and subsequent epochs in uniform steps. The curriculum for \(P_{Spk}\)  starts at 0 in the 1st epoch and increase to 0.14 in the 10th and subsequent epochs in uniform steps. The model is trained with Adam Optimizer with a batch size of 32 and an average sequence length of 30 words per batch. We use a learning rate of 1e-4 and train the model for 30 epochs on a machine with 8 GPUs.

\begin{figure}[t]
  \centering
  \includegraphics[width=1\columnwidth]{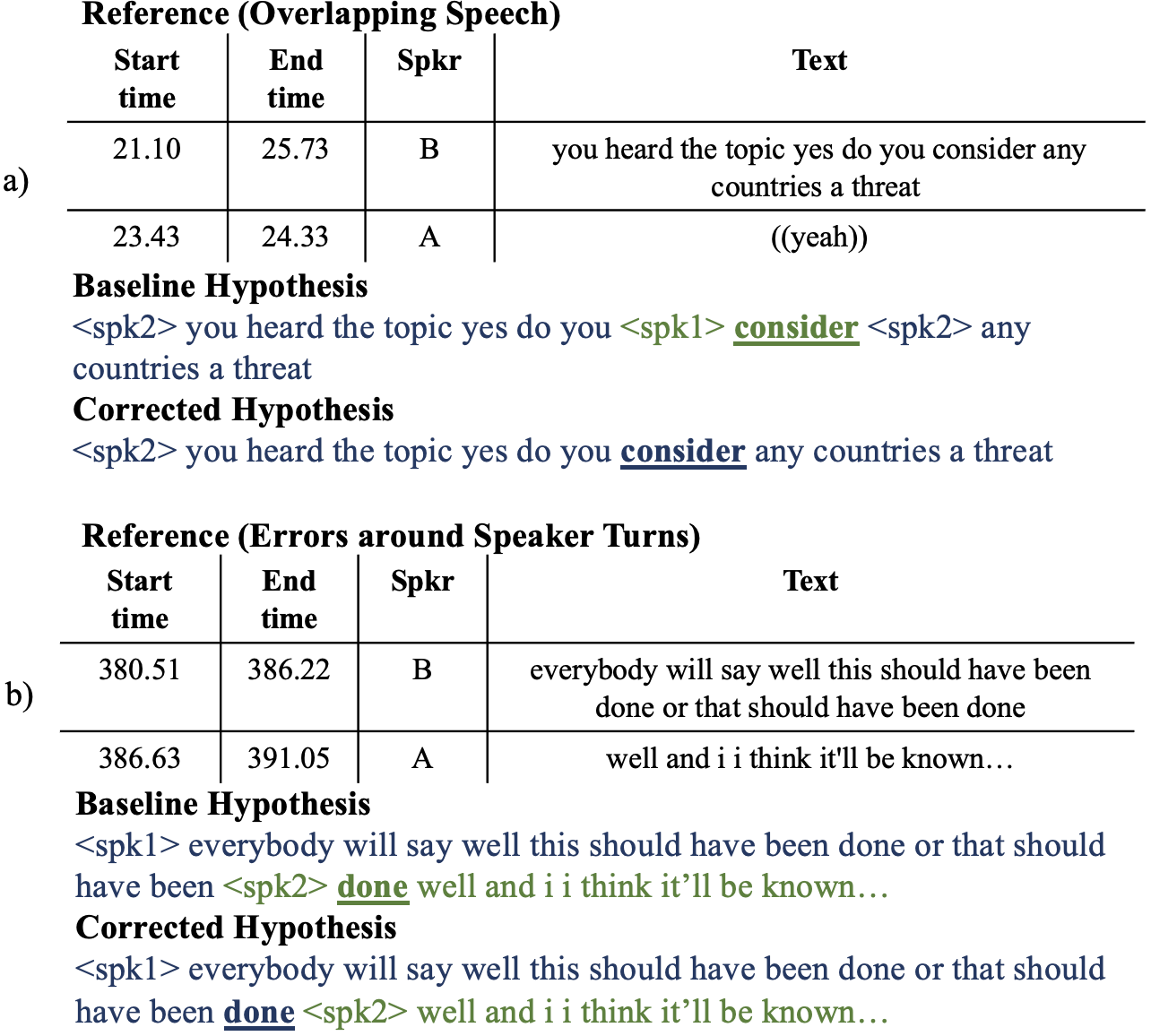}
  \vspace{-1\baselineskip}
  \caption{Correction Examples: a) Errors due to overlapping speech, b) Errors around speaker turns.}
  \label{fig:Correction Examples}
  \vspace{-1\baselineskip}
\end{figure}

We use the SEC as a 2nd pass post-processing step to the baseline SD-ASR system in Section 3.2. In order to determine the number of simulated errors needed to effectively train the lexical SEC to correct the speaker errors, we follow the error curricula mentioned in Section 2.4 and pick the checkpoint with \(P_{ASR}\), \(P_{Spk}\) that achieves the lowest WDER on the Fisher validation set. The values that achieve the best validation WDER are 0.1 for both \(P_{ASR}\) and \(P_{Spk}\). In addition to the training parameters, we also tune an inference parameter, the sliding window size as mentioned in Section 2.1 also on the Fisher validation set.

\begin{table*}[!ht]
  \caption{WDER of different models on Fisher test, RT03-CTS and CHAE dev, test sets. CHAE-109 is evaluated with and without fixing the number of speakers in 1st pass to 2. SimSEC: SEC model trained using simulated transcript errors, RealSEC: SEC trained/tuned on real paired data}
  \vspace{-0.2\baselineskip}
  \label{tab:data sampling table}
  \centering
    \begin{tabular}{P{50mm}P{12mm}P{10mm}P{10mm}P{10mm}P{10mm}P{15mm}P{15mm}} 
    \hline\hline
        \multirow{2}{*}{Model Type}   &\multirow{2}{*}{\makecell{Fisher\\Test}}  & \multicolumn{2}{c}{RT03-CTS} & \multicolumn{2}{c}{CHAE} & \multicolumn{2}{c}{CH-109} \\ \cline{3-8}
        & & Validation & Test & Validation & Test & \makecell{Known\\Spkrs} & \makecell{Unknown\\Spkrs} \\ \hline\hline
        Baseline (No Correction)  & 2.26  & 2.30 & 2.18 & 4.23 & 2.82 & 3.69 & 4.28 \\
        \makecell{SimSEC\_v1 (Base Roberta)} & 1.72 & 2.18 & 1.98 & 4.16 & 2.68 & 3.41 & 4.29 \\
        \makecell{SimSEC\_v2 (Tuned Roberta)} & 1.63  & 1.90 & 1.67 & 3.52 & 2.49 & 3.16 & 3.74 \\
       \makecell{RealSEC (flat-start Training)} & 1.53 & 1.73 & 1.58 & 3.31 & 2.30 & 2.98 & 3.57 \\ 
       \makecell{SimSEC\_v2 init + RealSEC Tuning} & 1.53 & 1.73 & 1.59 & 3.28 & 2.26 & 2.97 & 3.56 \\\hline
    \end{tabular}
    \vspace{-0.5\baselineskip}
\end{table*}

\begin{table}[t]
  \caption{WDER of models with different amounts of Synthetic/Real Data on Fisher test set.}
  \vspace{-0.2\baselineskip}
  \label{tab:data sampling table}
  \centering
    \begin{tabular}{P{20mm}P{25mm}P{20mm}} 
    \hline\hline
        Model Type   & \makecell{Fraction of \\Train Data}  & WDER \\ \hline\hline
        \multicolumn{2}{c}{Baseline (No Correction)} & 2.26 \\ \hline
        \multirow{4}{*}{\makecell{SimSEC\_v2}} & \multicolumn{1}{c}{0.2} & 1.73 \\
         & \multicolumn{1}{c}{0.4}   & 1.68 \\
         & \multicolumn{1}{c}{0.8}   & 1.65 \\
         & \multicolumn{1}{c}{1.0}   & 1.63 \\ \hline
        \multirow{4}{*}{\makecell{SimSEC\_v2 init + \\RealSEC Tuned}} & \multicolumn{1}{c}{0.2} & 1.58 \\
         & \multicolumn{1}{c}{0.4}   & 1.57 \\
         & \multicolumn{1}{c}{0.8}   & 1.54 \\
         & \multicolumn{1}{c}{1.0}   & 1.54 \\ \hline
    \end{tabular}
    \vspace{-1.6\baselineskip}
\end{table}

\subsection{Results}
We tune for the sliding window size on our Fisher validation subset and plot the WDER with the corresponding values as shown in Figure 2. From the plot, we see that WDER decreases as the window size increases up to 30 due to increased lexical context for the backbone LM as well as the corrector model. The WDER further increases beyond the window size of 30 likely due to the corrector model being trained with an average seq length of 30 and more sliding windows with greater than 2 speakers being bypassed with a larger window size. We have also tried training with larger average sequence lengths but that did not show any additional gains compared to the sequence length of 30 words. So, we use the sliding window size as 30 words for the remainder of the experiments. We also show some qualitative examples of the correction performance on the Fisher test set using the SEC model with the best sliding window size in Figure 3. Figure 3a shows that the correction model is able to effectively correct errors due to overlapping speech when the SD hypothesizes one of the overlapping speakers and ASR hypothesizes the words of the other speaker. The model is also effective in correcting the lexically implausible errors around speaker turns  which is one of the major error-prone scenario \cite{knox2012did} for SD systems as seen in Figure 3b.

The quantitative WDER improvements of the correction models on the held out validation and test sets are outlined in Table 1. We call the model trained on ground truth transcripts with simulated speaker, ASR errors as the "SimSEC" model. SimSEC\_v2 is the "SimSEC" model with a Fisher tuned backbone Roberta and trained with a custom curriculum as mentioned in Section 3.3 SimSEC\_v1 is similar to SimSEC\_v2 but with the Roberta-base as a backbone without any further fine-tuning. We evaluate SimSEC\_v1 to quantify the gains attributed to fine-tuning of the backbone LM on conversational datasets. "SimSEC init + RealSEC Tuning" model is the paired data tuned model initialized with SimSEC\_v2 and tuned using the 1st pass acoustic SD labels instead of the simulated speaker errors. RealSEC model is similar to "SimSEC\_v2 init + RealSEC Tuning" but is only trained by flat-starting the model on real paired data.

From Table 1, we can see that almost all of the corrector models produce considerable WDER gains over the Baseline SD-ASR reconciled system across all the datasets, except from SimSEC\_v1 on CH-109 with unknown speakers. It can be observed that tuning the backbone Roberta LM in SimSEC\_v2 can produce moderate WDER gains over the pretrained Roberta-base LM, especially on CHAE validation set and CH-109 with unknown speakers. The model trained on Paired data, either by tuning the SimSEC\_v2 model or by flat-start training (RealSEC) produces further gains over the models train with errors simulated (SimSEC\_v1 and SimSEC\_v2). The performance improvement of the models on CH-109 without fixing the speakers to 2 in the Clustering phase is comparatively limited due to hypothesising more than 2 speakers on few of the audio files leading to smaller average WDER gains. With the best model "SimSEC\_v2 init + RealSEC Tuning", we observe relative WDER gains in the range 15-30\% across all the datasets. 

To further analyze the importance of dataset sizes needed to train or tune the models, we perform an ablation by only using a fraction of the Fisher train data as shown in Table2. We evaluate The models SimSEC\_v2 and "SimSEC init + RealSEC Tuning" with different fractions of ground truth text and paired data respectively. We see that the WDER of the SimSEC\_v2 model and "SimSEC init + RealSEC Tuning" model only improves moderately and saturates at a point as the amount of text data and paired data increases respectively. This shows that the corrector model can be trained purely on small amounts of ground truth transcripts by simulating speaker, ASR errors and can also be fine-tuned on a small amount of paired data to achieve significant WDER gains.  
\section{Conclusion and Future Work}
In this work, we propose a novel Speaker Error Corrector (SEC) to correct word-level speaker label errors from a conventional audio-only speaker diarization system. We achieve this using a language model over the ASR transcriptions to correct the speaker labels. The proposed lexical SEC can be trained effectively using only text data by simulating speaker errors without the need for any paired audio-text data. A small amount of paired data can further improve model performance, leading to overall relative reduction of WDER by over 15\% across three telephony datasets. The proposed SEC framework is also lightweight and is easy to integrate as a post-processing module over existing systems.

One limitation of our current work is that it has been applied only to conversations in English. One future work can include training a multi-lingual SEC to make the system language-agnostic. To increase the robustness of this approach, in addition to the first-pass SD labels, we can leverage additional complementary acoustic cues to further improve the performance. Also, the current SEC model can only handle 2 speakers in a sliding window, which we plan to generalize to handle more number of speakers. We will also explore leveraging large generative models to synthesize conversational transcripts across multiple domains using curated prompts \cite{chen2023places}.

\bibliographystyle{ieeetr}
{\eightpt
\bibliography{citations}}

\end{document}